\documentclass[preprintnumbers,amsmath,amssymb]{revtex4}
\usepackage{graphicx}
\usepackage{dcolumn}
\usepackage{bm}
\begin{document}
\def\oppropto{\mathop{\propto}} 
\def\opsimeq{\mathop{\simeq}}
\def\opoverderline{\mathop{\overline}}
\def\operarrow{\mathop{\longrightarrow}}
\def\opsim{\mathop{\sim}} 

\title {Delocalization transition of the selective interface model: \\
distribution of pseudo-critical temperatures}
\author{ C\'ecile Monthus and Thomas Garel }
 \affiliation{Service de Physique Th\'{e}orique, CEA/DSM/SPhT\\
Unit\'e de recherche associ\'ee au CNRS\\
91191 Gif-sur-Yvette cedex, France}

\begin{abstract}
According to recent progress in the finite size scaling theory of
critical disordered systems, the nature of the phase transition is
reflected in the distribution of pseudo-critical temperatures
$T_c(i,L)$ over the ensemble of samples $(i)$ of size $L$. In this
paper, we apply this analysis to 
the delocalization transition of an heteropolymeric chain at a selective
fluid-fluid interface. The width $\Delta T_c(L)$ and the shift
$[T_c(\infty)-T_c^{av}(L)]$ are found to decay with the same exponent
$L^{-1/\nu_{R}}$, where $1/\nu_{R} \sim 0.26$. The distribution of
pseudo-critical temperatures $T_c(i,L)$ is clearly asymmetric, and is
well fitted by a generalized Gumbel distribution of parameter $m 
\sim 3$. We also consider the free energy distribution, which can also
be fitted by a generalized Gumbel distribution with a temperature
dependent parameter, of order $m \sim 0.7$ in the critical region. Finally,
the disorder averaged number of contacts with the interface scales at
$T_c$ like $L^{\rho}$ with $\rho \sim  0.26 \sim 1/\nu_R $. 

\bigskip


\end{abstract}
\maketitle

\section{Introduction}
Heteropolymers containing both hydrophobic and hydrophilic components
are of particular interest in biology \cite{review1}.
In a polar solvent, these heteropolymers
prefer conformations where the hydrophilic components
are in contact with the polar solvent, whereas hydrophobic
components avoid contacts with the solvent.
The behavior of such heteropolymers in the presence of an interface 
separating two selective solvents, one favorable to the hydrophobic
components and the other to the hydrophilic components,
is less obvious, and has been much studied recently \cite{Fleer,protein1}.
In the initial work of ref. \cite{garel},
a model was proposed and studied via Imry-Ma arguments, 
an analysis of the replica Hamiltonian and numerics : it was found
that a chain with a symmetric distribution of hydrophobic/hydrophilic
 components is always localized around the interface
at any temperature (in the thermodynamic limit), whereas a chain
 with a dissymmetric distribution of hydrophobic/hydrophilic
components presents a phase transition separating 
a localized phase at low temperature from a delocalized phase
into the most favorable solvent at high temperature.
 Experimentally, the presence of
copolymers was found to stabilize the interface
between the two immiscible solvents \cite{experiments},
since the localization of the heteropolymers at the interface
reduces the surface tension.
By now, the predictions of ref. \cite{garel} have been confirmed 
in the physics community by various approaches 
including molecular dynamics simulations \cite{yeung},
Monte Carlo studies \cite{sommer,Chen},
variational methods for the replica Hamiltonian
\cite{stepanow} \cite{maritanreplica}, exact bounds for the
free-energy \cite{maritanbounds}, and dynamical approach
\cite{Gan_Bre}. A real space renormalization group study, based on
rare events, was proposed in \cite{CM2000}.
Mathematicians have also been interested in this model. The
localization at all temperatures for the symmetric case was proven in
ref. \cite{sinai,albeverio}. In the dissymmetric case, the
existence of a transition line in temperature vs dissymmetry plane was
proven in ref. \cite{bolthausen,biskup}. Recently, important progresses 
were made in various aspects of the phase transition
\cite{GT1,GT2,GT3,GT4}.  

We study here the same model using the Wiseman-Domany approach to finite
size scaling in disordered systems
\cite{domany95,AH,Paz1,domany,AHW,Paz2}. The main outcome of this approach
can be summarized as follows.
To each disordered sample $(i)$ of size $L$, one should first associate
a pseudo-critical temperature $T_c(i,L)$, defined for instance
in magnetic systems as the temperature where the susceptibility is
maximum \cite{domany95,Paz1,domany,Paz2}. 
The disorder averaged pseudo-critical temperature $T_c^{av}(L) \equiv
\overline{T_c(i,L)}$ satisfies 
\begin{equation}
T_c^{av}(L)- T_c(\infty) \sim L^{-1/\nu_{R}}
\label{meantc}
\end{equation}
where $\nu_R$ is the correlation length
exponent. Eq. (\ref{meantc}) generalizes the analogous relation for
pure systems 
\begin{equation}
\label{puretc}
T_c^{pure}(L) - T_c(\infty) \sim L^{-1/\nu_{P}}
\end{equation}
The nature of the disordered critical point then depends on the
width 
\begin{equation}
\label{variance}
\Delta T_c(L) \equiv \sqrt{Var [T_c(i,L)]}
=\sqrt{\overline{T_c^2(i,L)}-\left(\overline{T_c(i,L)}\right)^2}
\end{equation}
of the distribution of pseudo-critical temperatures. When the
disorder is irrelevant, the fluctuations
of these pseudo-critical temperatures obey
the scaling of a central limit theorem as in the Harris argument : 
\begin{equation}
\Delta T_c(L) \sim L^{-d/2} \ \ \hbox{ for  irrelevant  disorder }
\label{deltatcirrelevant}
\end{equation}
This behavior was first believed to hold in general \cite{domany95,Paz1}, 
but was later shown to be wrong in the case of random fixed points :
in \cite{AH,domany}, it was argued that
at a random critical point, one has instead 
\begin{equation}
\Delta T_c(L) \sim L^{-1/\nu_{R}} \ \ \hbox{ for  random critical points }
\label{deltatcrelevant}
\end{equation}
i.e. the scaling is the same as the $L$-dependent shift of the averaged 
pseudo-critical temperature (\ref{meantc}).
The fact that these two temperature scales remain the same
is then an essential property of random fixed points
that leads to the lack of self-averaging for observables at criticality
\cite{AH,domany}. 

In this paper, we apply this finite size scaling analysis to the case
of the polymer chain at an interface between two selective fluids.
The paper is organized as follows. The model is defined in Section
\ref{model}. The definition of
pseudo-critical temperatures, and their 
distribution is presented in Section \ref{distritc}. The distribution
of the free energy difference between delocalized and localized phases
is studied in Section \ref{distrifree}. Finally, in Section
\ref{contact}, we consider the averaged number of contacts with the
interface. The numerical implementation through the Fixman-Freire
procedure is postponed to the Appendix.

\section{Model and notations}
\label{model}

\subsection{Definition of the model } 

The model we consider is defined by the partition function
 \begin{eqnarray}
\label{definition}
Z(L) = \sum_{ \{z_{\alpha}\} }
exp \left(\beta \sum_{\alpha} q_{\alpha} {\rm sgn}z_{\alpha} \right)
\end{eqnarray}
with inverse temperature $\beta =1/T$. The sum in (\ref{definition}) is
over all random walks $\{z_{\alpha}\}$ of $L$ steps, 
with increments $z_{{\alpha}+1}-z_{\alpha}=\pm 1$ and (bound-bound)
boundary conditions $z_1=0=z_L$. In eq. (\ref{definition}), the
$\{q_{\alpha}\}$ represent quenched random charges, and with our 
convention, monomers with positive charges prefer to be in the upper
fluid (${\rm sgn }z >0$). Furthermore, we have taken
$q_{2{\alpha}+1}=0$ for all $(\alpha)$, so that there is no frustration at 
zero temperature (each monomer $z_{2{\alpha}}$ being in its preferred
solvent). Even charges $q_{2{\alpha}}$ are
random and drawn from the Gaussian distribution with mean value $q_0$
($q_0>0$) 
 \begin{eqnarray}
\label{gauss}
P(q_{2{\alpha}})= \frac{1}{ \sqrt{2 \pi \Delta^2} } e^{-
\frac{(q_{2{\alpha}}-q_0)^2} {2 \Delta^2} }
\end{eqnarray}

From previous work, model (\ref{definition}) is expected to undergo a
phase transition at a critical temperature $T_c \simeq
O(\Delta^2/q_0)$ between a localized phase and a delocalized phase in
the upper fluid.

\subsection{Recursion relations for the partition function } 

Let $Z(\alpha)$ be the partition function of the partial chain
($1,2,...\alpha$), with monomer ($\alpha$)
sitting at the interface ($z_{\alpha}=0$). Since the last loop before
monomer $(\alpha$) can be in the upper ($+$) or in the lower ($-$)
fluid, we may write
 \begin{eqnarray}
Z(\alpha) = Z^{+}(\alpha) + Z^{-}(\alpha)
\label{ztot}
\end{eqnarray}
The partition functions
$Z^{\pm}(\alpha)$ satisfy the recursions
 \begin{eqnarray}
Z^{\pm}(\alpha) = \sum_{\alpha'=1,3,..}^{\alpha-2} 
exp \left(\pm \beta \sum_{i=\alpha'+1}^{\alpha-1} q_i  \right) \ 
{\cal N} (\alpha', \alpha) \left[Z^{+}(\alpha') +
Z^{-}(\alpha') \right] 
\label{rec1}
\end{eqnarray}
where ${\cal N} (\alpha', \alpha)$ represents the number of random
walks of $(\alpha-\alpha')$ steps going from $z_{\alpha'}=0$ to
$z_{\alpha}=0$, without touching the interface at $z=0$ in between.
For the critical properties of the delocalization transition, the only
important property of ${\cal N} (\alpha', \alpha)$ is its behavior at
large separation 
 \begin{eqnarray}
{\cal N} (\alpha', \alpha) \opsimeq_{\alpha-\alpha' \gg 1} 
 \sigma \frac{2^{\alpha-\alpha'}}{(\alpha-\alpha')^{3/2}}
\end{eqnarray}
where $\sigma$ is a constant of order $1$.
In our numerical study, we have chosen to take 
the following form
 \begin{eqnarray}
{\cal N} (\alpha', \alpha) = 
 \sigma \frac{2^{\alpha-\alpha'}}{(\alpha-\alpha'-1)^{3/2}}
\end{eqnarray}
for all $(\alpha -\alpha')$.
If one forgets the boundary conditions at $(1,\alpha)$, the partition
function characterizing the delocalized phase in the $(+)$ solvent
($q_0>0$) would simply be for each sequence of charges
 \begin{eqnarray}
Z^{deloc}(\alpha) = 2^{\alpha-1} e^{\beta V(\alpha) }
\ \ \  { \rm with} \ \ \ 
 V(\alpha) \equiv  \sum_{\alpha'=2}^{\alpha} q_i
\label{zdeloc}
\end{eqnarray}
It is convenient to introduce the ratio
 \begin{eqnarray}
{\tilde Z}(\alpha)
 \equiv \frac{Z(\alpha)}{Z^{deloc}(\alpha)}
= {\tilde Z}^{+}(\alpha) + {\tilde Z}^{-}(\alpha)
\label{defztilde}
\end{eqnarray}
where $
{\tilde Z}^{\pm}(\alpha)
 \equiv Z^{\pm}(\alpha)/Z^{deloc}(\alpha)$
satisfy the recursion relations (Eq. \ref{rec1})
 \begin{eqnarray}
{\tilde Z}^{+}(\alpha) && = \sum_{\alpha'=1,3,..}^{\alpha-2} 
  \frac{  \sigma }{(\alpha-\alpha'-1)^{3/2}} \left[{\tilde Z}^{+}(\alpha') +
{\tilde Z}^{-}(\alpha') \right] \nonumber \\
{\tilde Z}^{-}(\alpha) && = \sum_{\alpha'=1,3,..}^{\alpha-2} 
e^{ -2 \beta \left(  V(\alpha-1) -  V(\alpha')  \right) }
  \frac{  \sigma }{(\alpha-\alpha'-1)^{3/2}} \left[{\tilde Z}^{+}(\alpha') +
{\tilde Z}^{-}(\alpha')\right]
\label{rec2}
\end{eqnarray}

The numerical iteration of Eqs (\ref{rec2}) takes a CPU time of
order $O(L^2)$ for a chain of length $L$, but it is possible
to obtain a CPU time of order $O(L)$ via a Fixman-Freire
scheme \cite{Fix_Fre}. Originally proposed in the context of the
Poland-Scheraga model of DNA denaturation, this scheme can be adapted
to the interface model, as explained in the Appendix.

\subsection{Numerical parameters } 
 We have taken $q_0=1$ and  $\Delta=10$ in the Gaussian distibution of
eq (\ref{gauss}). The data we present correspond to various sizes $L$
with corresponding numbers $n_s(L)$ of independent samples.
Unless otherwise stated, we have considered the 
following sizes going from $L=2 \cdot 10^3$ to $L=2048 \cdot 10^3 $, 
with respectively $n_s(L=16 \cdot 10^3)=480 \cdot 10^3 $ to $n_s(L=2048 \cdot
10^3)= 15 \cdot 10^3$. More precisely, we consider
\begin{eqnarray}
\label{sizesandnseed}
 \frac{L}{1000} && =  16, 32, 64, 128 , 256, 512, 1024, 2048   \\
  \frac{n_s(L)}{1000} && =  480, 240, 120, 60 , 30, 15, 30 ,15 
\end{eqnarray}

\section{Distribution of Pseudo critical temperatures}
\label{distritc}

In a previous paper \cite{PSdisorder}, we have discussed two
possible definitions for the pseudo-critical temperatures for
Poland-Scheraga models. Here we use the definition based on the free
energy, as we now explain.

\subsection{Definition of a sample dependent $T_c(i,L)$}

For each sample $(i)$ of length $L$, we may define a pseudo-critical
temperature $T_c(i,L)$ as the temperature where the free energy
$F^{(i)}(L,T) \equiv -T \ln Z^{(i)}(L,T)$
 reaches the delocalized value
 $F_{deloc}^{(i)}(L,T)=-T (L-1) \ln 2 -V(L)$ (Eq. \ref{zdeloc}),
i.e. $T_c(i,L)$ is the solution of the equation
 \begin{eqnarray}
\label{fre}
F^{(i)}(L,T)+T (L-1) \ln 2+V(L) =0
\end{eqnarray}

As explained in details in our previous study \cite{PSdisorder},
this definition of pseudo-critical critical temperatures, together
with bound-bound boundary conditions, introduces logarithmic
correction in the convergence towards
$T_c(\infty)$. Eq. (\ref{meantc}) is accordingly replaced by 
\begin{equation}
T_c^{av}(L)- T_c(\infty) \sim \left( \frac{\ln L}{ L} \right)^{1/\nu_{R}}
\label{meantc2}
\end{equation}

\subsection{Scaling form of the probability distribution}

\begin{figure}[htbp]
\includegraphics[height=6cm]{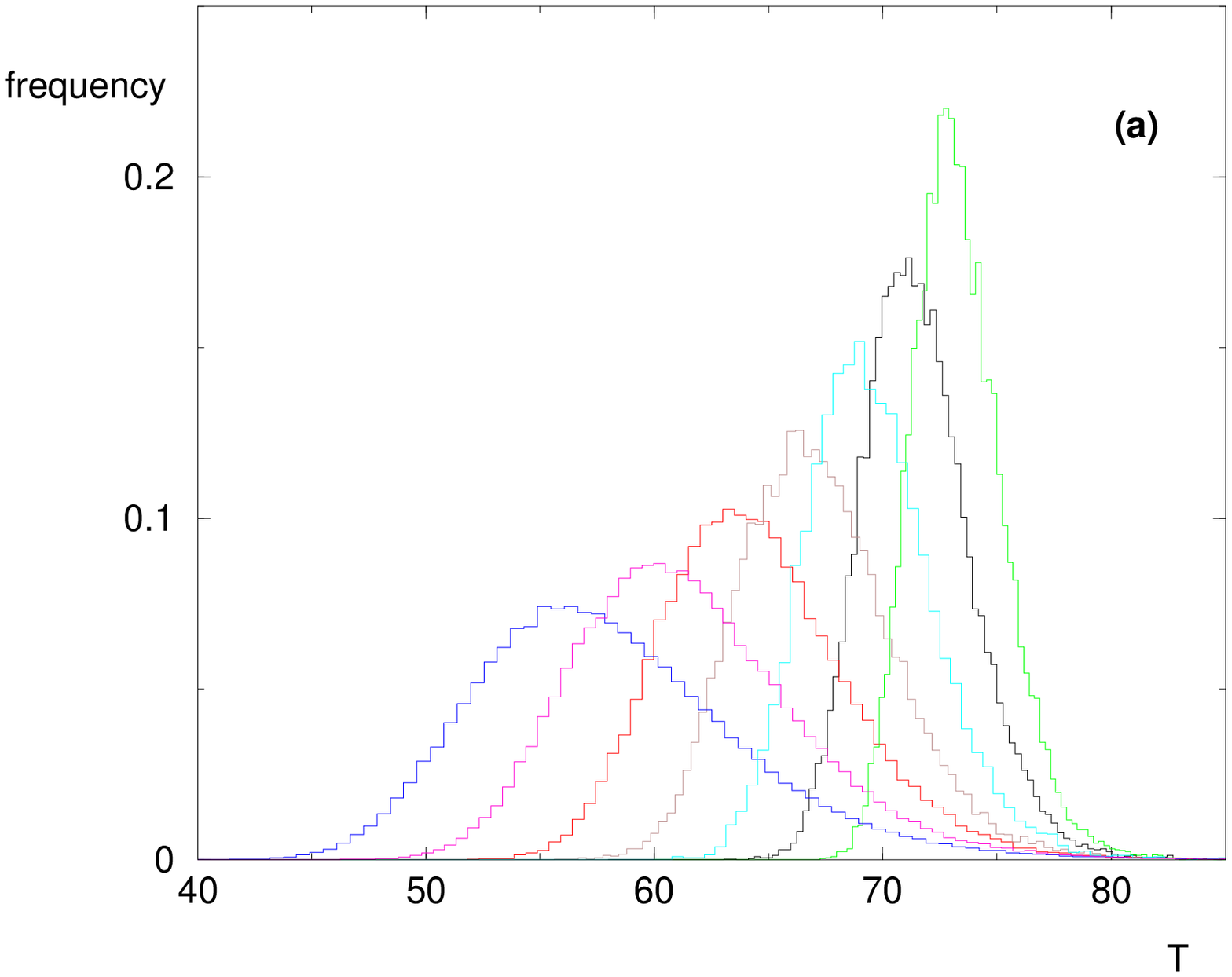}
\hspace{1cm}
\includegraphics[height=6cm]{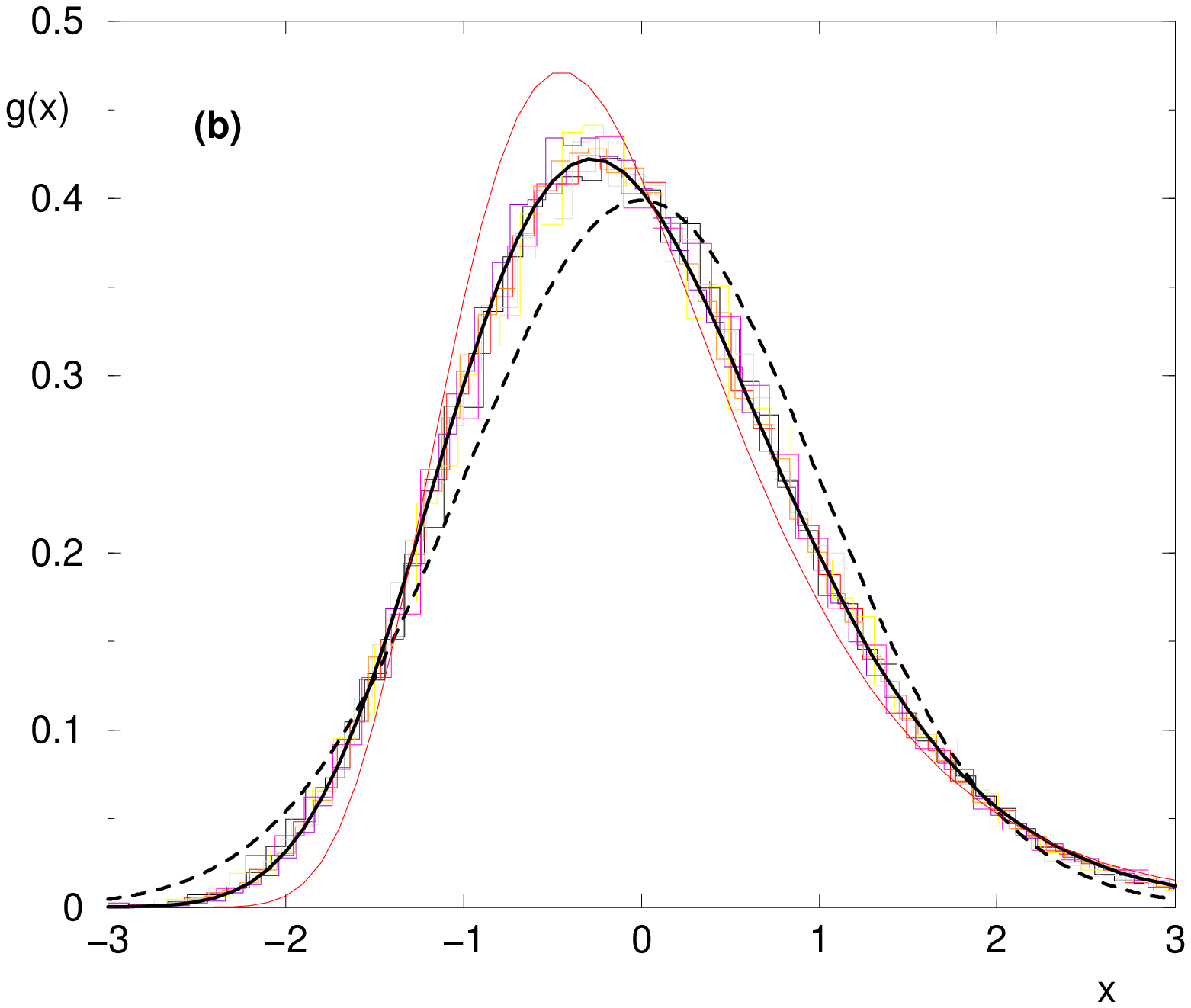}
\caption{(a) Distribution of critical temperatures for
($L/10^3=32,64,128,256,512,1024,2048 $)
(b) Same data after the rescaling of eq. (\ref{rescalinghistotc}): fit
with the generalized Gumbel distribution $G_3(-x)$ (thick line). For
comparison, we also show the usual Gumbel $m=1$ (thin line) and
Gaussian $m=\infty$ (dashed line) 
distributions.} 
\label{gumbel1}
\end{figure}

Our data for the distribution of pseudo-critical temperatures (see
Figure \ref{gumbel1}) follow the scaling form
\begin{equation}
P_L(T_c(i,L)) \simeq  \frac{1}{ \Delta T_c(L)} \  g \left( x= \frac{
T_c(i,L) -T_c^{av}(L)}{ \Delta T_c(L) }  \right) 
\label{rescalinghistotc}
\end{equation}
where the scaling distribution $g(x)$ (normalized with $<x>=0$ and
$<x^2>=1$) is well fitted by the one parameter generalized 
Gumbel distribution $G_m(-x)$
\begin{equation}
\label{Gumb1}
g(x)=  G_m(-x) \equiv \frac{m^m}{\Gamma(m)}\left( e^{-x-e^{-x}}\right)^m
\end{equation}
with parameter $m \sim 3$. On Fig. \ref{gumbel1}(b), we show the fit of our
data with $G_3(-x)$, and plot for comparison the usual Gumbel ($m=1$)
and Gaussian ($m \to \infty$) distributions.

\subsection{Scaling properties of the shift and of the width}

\begin{figure}[htbp]
\includegraphics[height=6cm]{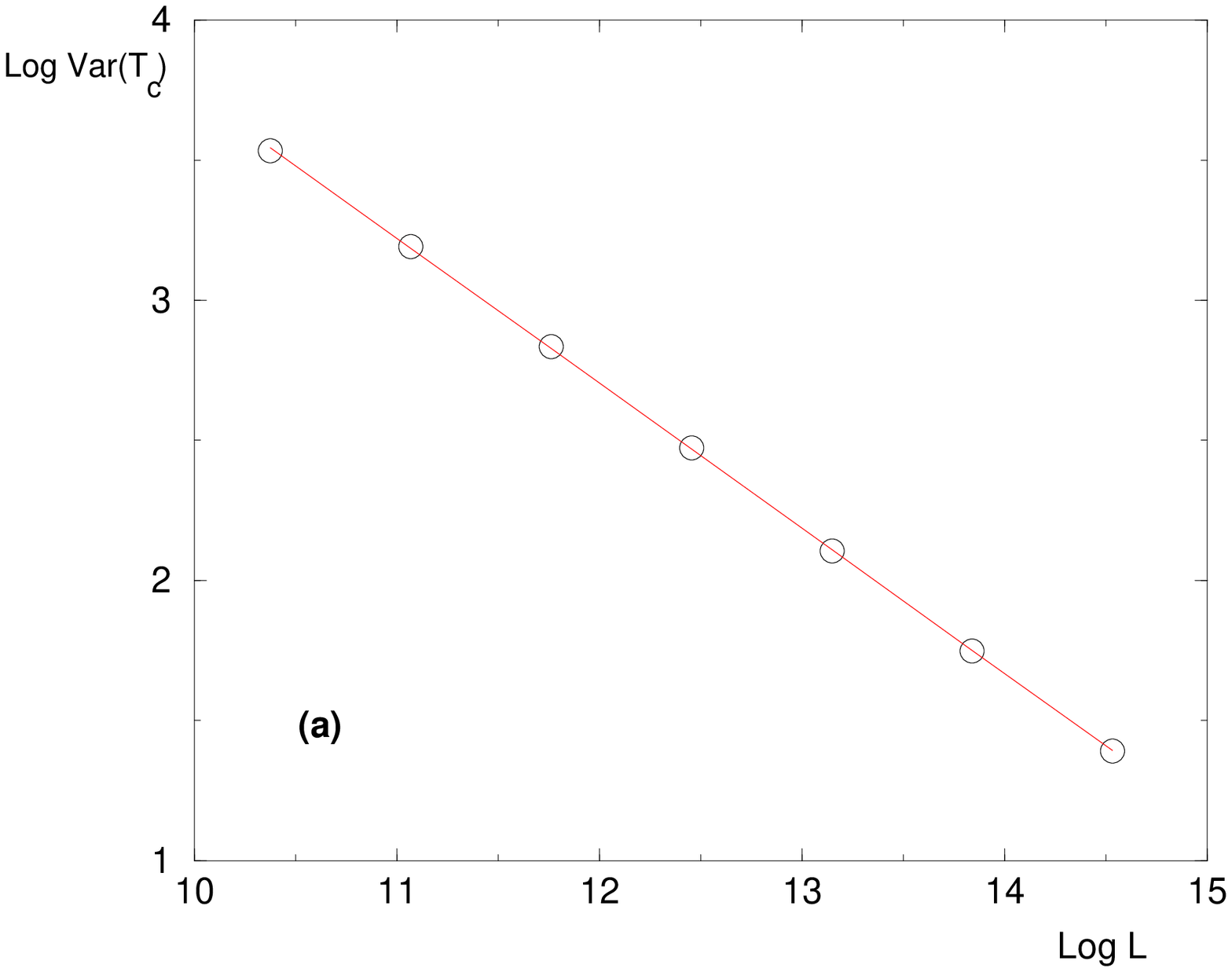}
\hspace{1cm}
\includegraphics[height=6cm]{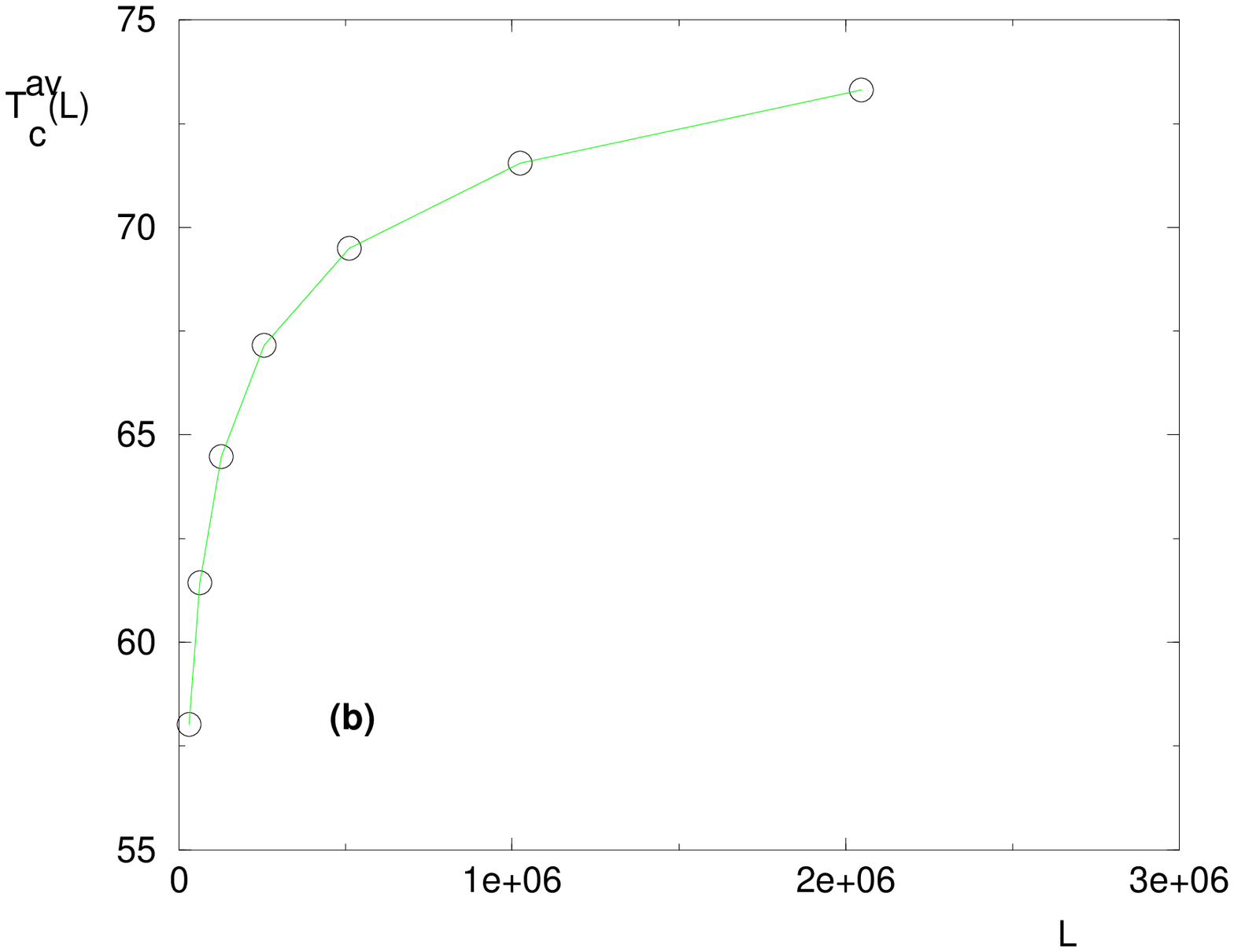}
\caption{(a) Log-Log plot of the variance ${Var [T_c(i,L)]}=(\Delta
T_c(L))^2$ for $L/10^3=32,64,128,256,512,1024,2048 $ ($\bigcirc$) with
the fit of equation (\ref{reswidth})  (b) The average $T_c^{av}(L)$
for the same sizes, with the fit of equation (\ref{resav}) .} 
\label{45loglog}
\end{figure}

We now consider the behavior of the width $\Delta T_c(L)$ and of the
average $T_c^{av}(L)$ of the distribution (\ref{rescalinghistotc}), as
$L$ varies. For the width $\Delta T_c(L)$, we have fitted our data
with the following power law (see Fig \ref{45loglog}(a))
\begin{eqnarray}
\label{reswidth}
\Delta T_c(L)  \simeq 86  \left(\frac{1}{
L}\right)^{0.26} 
\end{eqnarray}

For the average $T_c^{av}(L)$ (see Fig \ref{45loglog}(b)), our data
are well fitted by the generalized form of eq. (\ref{meantc2})
\begin{eqnarray}
\label{resav}
T_c^{av }(L)  \simeq 83.8
-246  \left(\frac{\ln (L/289)}{L} \right)^{0.26}
\end{eqnarray}
The value $T_c(\infty) \simeq 83.8=0.838 \frac{\Delta^2}{q_0}$ is in
agreement with the numerical estimate of ref \cite{GT3}. The result
for the exponent in eqs (\ref{reswidth}) and (\ref{resav}) indicate
that the transition can be described as a random critical point with
a single correlation exponent 
\begin{equation}
\frac{1}{\nu_R} \simeq 0.26
\end{equation}
A similar value has been observed in numerical simulations by the
authors of ref. \cite{GT3} (private commnunication ).

Our data rule out the possibility of an infinite order transition
based on rare negatively charged sequences
\cite{CM2000}. The present results suggest that
the excursions in the unfavorable fluid that are important for
the transition, are of finite length. 

\section{Free energy distribution}
\label{distrifree}

Since the distribution of pseudo critical temperature is clearly
linked to the statistical properties of the free energy, we have
studied the probability distribution of the free energy difference
$\tilde F^{(i)}(L,T)$ defined by (see eq. \ref{fre}) 
\begin{equation}
\label{freedist}
\tilde F^{(i)}(L,T) \equiv F^{(i)}(L,T)-F_{deloc}^{(i)}(L,T)=F^{(i)}(L,T)+T
(L-1) \ln 2+V(L) 
\end{equation}

\begin{figure}[htbp]
\includegraphics[height=6cm]{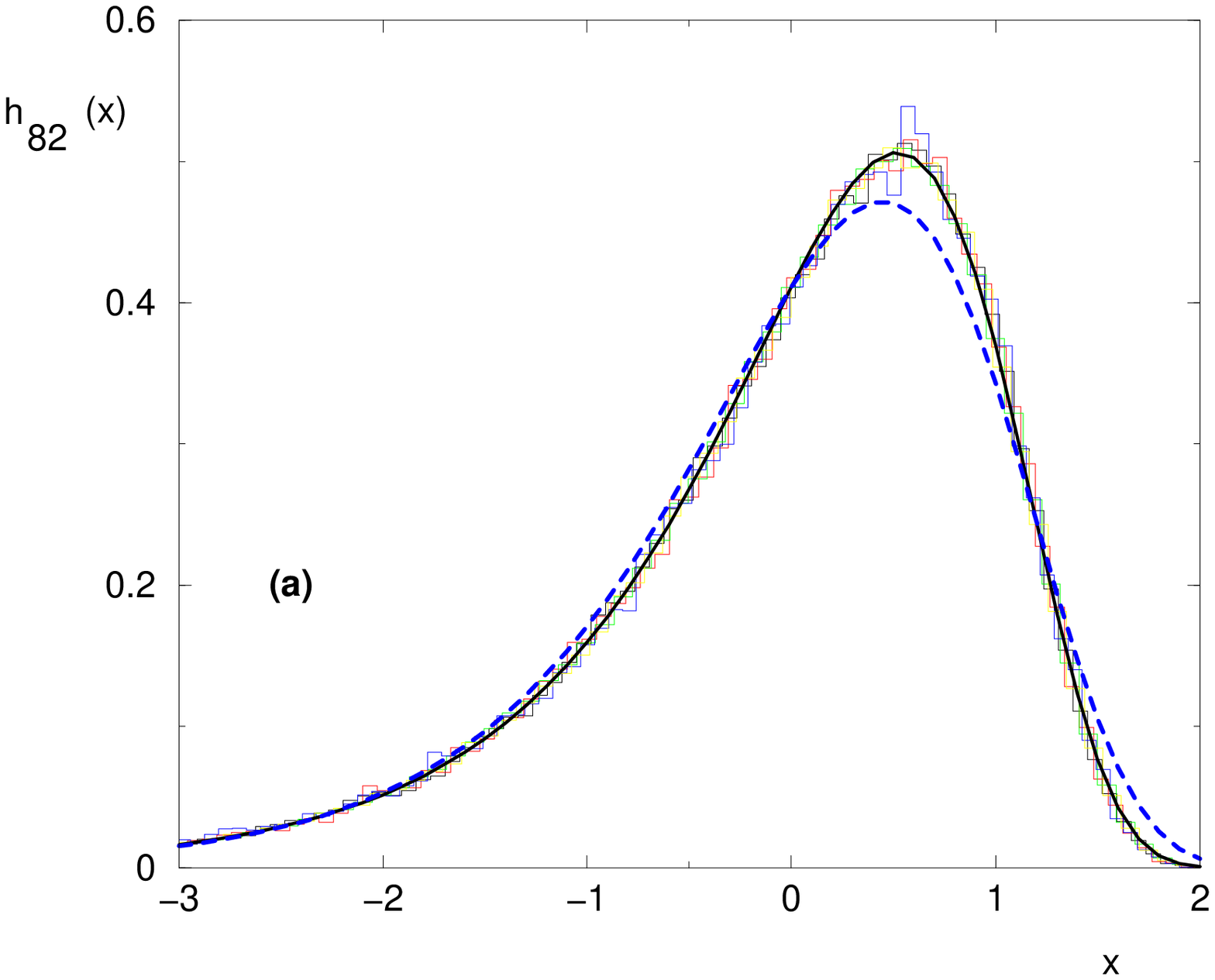}
\hspace{1cm}
\includegraphics[height=6cm]{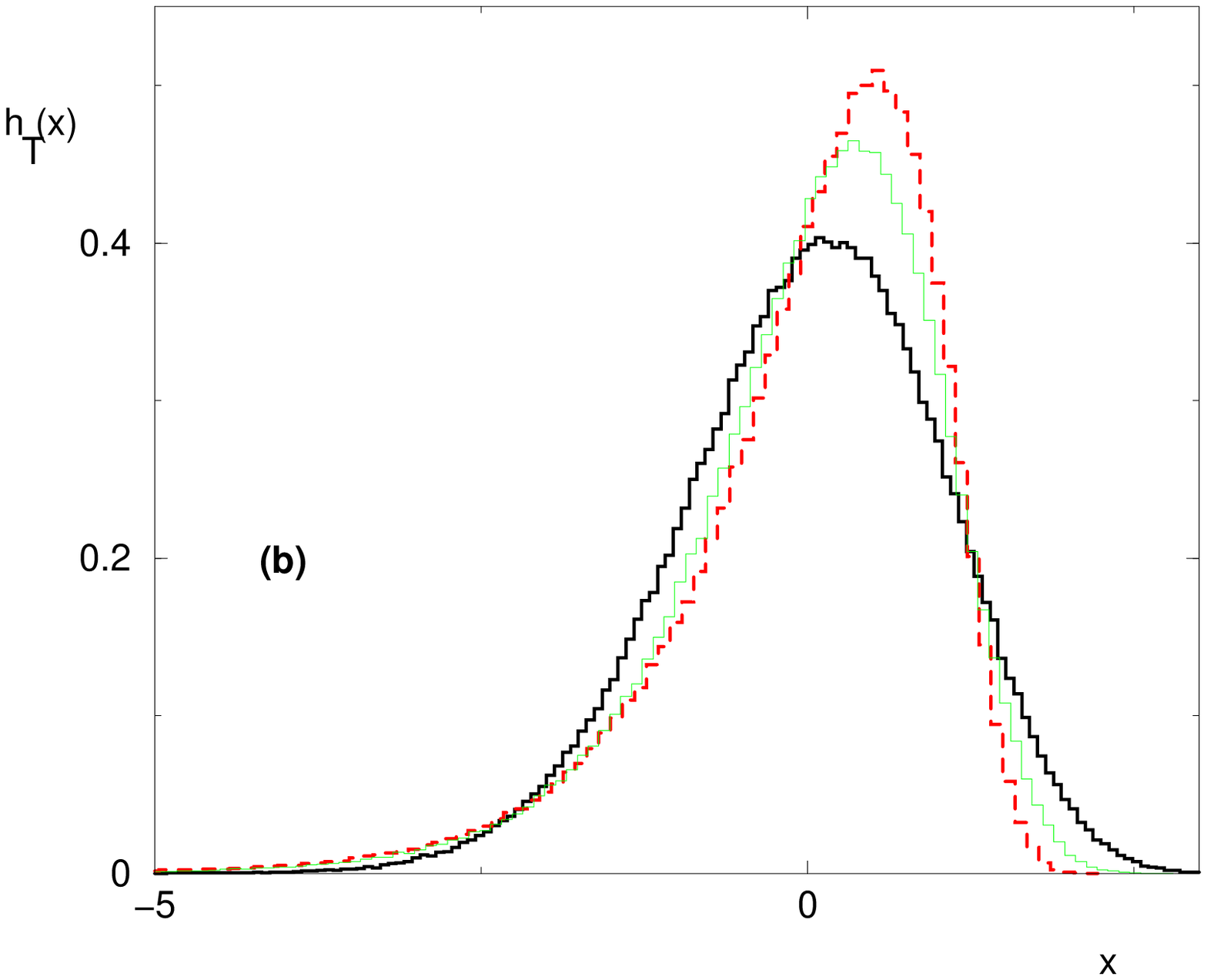}
\caption{ (a) Fit of the rescaled free energy distribution $h_T(x)$
(eq.(\ref{rescalinghistofre})) at $T=82$ for
$L/10^3=32,64,128,256,512$ by the generalized Gumbel distribution
$G_m(x)$ with $m=0.66$ (thick line). The usual Gumbel distribution
($m=1$) is also shown for comparison (dashed line)
(b) The function $h_T(x)$ for $L/10^3=32$ at $T=40$ (thick solid
line), $T=82 \sim T_c$ (dashed line) and $T=130$ (thin solid
line). Fitting with the generalized Gumbel distribution $G_m(x)$ leads
to a non monotonous variation of $m$ w.r.t the temperature $T$. The
minimal value of $m$ is obtained in the critical region.}
\label{histofree}
\end{figure}

Defining the average $\tilde F^{(av)}(L,T)$  and the width 
$\Delta \tilde F(L,T)$, we write the scaled probability distribution
as 

\begin{equation}
P(\tilde F^{(i)}(L,T)) \simeq  \frac{1}{\Delta \tilde F(L,T)} \ h_T
\left( x= \frac{ \tilde F^{(i)}(L,T)-\tilde F^{(av)}(L,T)}{ \Delta
\tilde F(L,T)  }  \right) 
\label{rescalinghistofre}
\end{equation}
where the function $h_T(x)$ is independent of $L$, but depends on the
temperature $T$. We show in Figure \ref{histofree} the function $h_T(x)$
for $T=40$ (deep in the localized phase), $T=82$ (critical region) and
$T=130$ (deep in the delocalized phase).

It turns out that a good one parameter fit is again obtained with the
generalized Gumbel distribution
\begin{equation}
\label{Gumb2}
h_T(x)=  G_m(x) \equiv \frac{m^m}{\Gamma(m)}\left( e^{x-e^{x}}\right)^m
\end{equation}
where the effective parameter $m$ depends on the temperature $T$. We now
characterize the behaviour at low and high temperatures, as well as in
the critical region.

\subsection{The localized phase}

At $T \ll T_c$, the sample dependent free energy $\tilde
F^{(i)}(L,T)$ can be fitted by 
\begin{equation}
\label{localized}
\tilde F_i(L,T)=Lf_{\infty}(T)+\sigma_T \sqrt{L} \ u_i +O(1)
\end{equation}
where $u_i$ is a Gaussian random
variable (which corresponds to the Gumbel parameter $m \to
\infty$). Eq. (\ref{localized}) is indeed expected from the existence
of an extensive number of contacts with the interface: fluctuations
are then of order $\sqrt{L}$ and Gaussian distributed.

\subsection{The delocalized phase}

At $T \gg T_c$ the sample dependent free energy $\tilde F_i(L,T)$ can
be fitted by
\begin{equation}
\label{delocalized}
\tilde F_i(L,T)=\frac{3}{2}T \ {\rm Ln}\frac{L}{l_0} +v_i
\end{equation}
where $l_0$ is a constant, and $v_i$ is a random
variable distributed with $G_m(x)$ and $m \sim 1.3$ (see eq. (\ref{Gumb2}).
The first term on the r.h.s of eq.(\ref{delocalized}) comes from the
bound-bound boundary conditions, while the random term of order $O(1)$
reflects rare contacts with the interface.

\subsection{In the critical region}

In the critical region, the average free energy $\tilde F^{(av)}(L,T)$
behaves as
\begin{equation}
\tilde F^{(av)}(L,T)=Lf_{\infty}(T)+f_1(T) \ {\rm Ln}\frac{L}{l_0}
\end{equation}
where $f_1$ is a coefficient which weakly depends on temperature in
the critical region. The thermodynamic free energy density
$f_{\infty}(T)$ is then found to go to zero between $T=82$ and $T=83$,
in good agreement with the value $T_c \simeq 83.8$ obtained above.
Concerning the rescaled distribution $h_T(x)$, we observe that the
parameter $m$ remains  at a minimum value of order $0.7$ in a large
temperature range $75<T<85$. This large crossover region is clearly
linked to the small value of $1/\nu_R \sim 0.26$.

\section{Contacts}
\label{contact}
We have also considered the disorder averaged number of contacts $N_0$
with the interface in the critical region, and we numerically obtain
\begin{equation}
N_0(T_c,L) =\sum_{\alpha=1}^L 
\overline{<\delta(z_{\alpha},0)>_{T_c}} \sim L^{\rho} \ , \
\rho \sim 0.26 
\end{equation}
It is interesting to compare this result with the behavior
of the internal energy difference $E_0(T,L)$ between the localized and
delocalized phases. A simple scaling argument based on the
hyperscaling relation $\alpha=2-\nu_R$ yields the following behavior
for the energy $E_0(T_c,L) \sim L^{1/\nu_R}$ at criticality. The fact
that the exponent $\rho$ is close to the exponent $1/\nu_R$ points
towards the same scaling for the energy and number of excursions in
the unfavorable fluid : this means that, at criticality, the
excursions in the lower unfavorable fluid are of finite length. 

Finally, our data on
the number of contacts $N_0(T,L)$ yield an estimate of the
critical temperature $82.5<T_c<83$, which is again compatible
with the above mentionned values.

\section{Conclusion}

We have studied the probability distribution finctions (PDF) of the
pseudo-critical temperatures $T_c(i,L)$ and of the free energy $\tilde
F_i(L,T)$ for the localization transition of an heteropolymer at a
selective fluid-fluid interface. We have obtained an estimate of the
critical temperature $T_c \sim 0.83 \frac{\Delta^2}{q_0}$, in
agreement with \cite{GT3}. Both the width $\Delta T_c(L)$ and the shift
$[T_c(\infty)-T_c^{av}(L)]$ are found to decay with the same exponent
$L^{-1/\nu_{R}}$ where $1/\nu_{R} \sim 0.26$. These results show
that the transition is not due to rare events, such as large
excursions in the unfavorable fluid \cite{CM2000}.

Concerning the form of the PDF's, we find here
asymmetric distributions, which can fitted with generalized Gumbel
distributions $G_m(x)$. This is in contrast with the wetting
and Poland-Scheraga transitions \cite{PSdisorder}, where these PDF's
were found to be Gaussian. Generalized Gumbel distributions have been
recently used in various contexts
\cite{Bramwell,BBW,Palassini,Rosso,Bertin}. As discussed in
ref. \cite{Bramwell}, it is empirically known that PDF's with the same
first four moments approximately coincide over the range of a few
standard deviation which is precisely the range of numerical -or
experimental- data. So generalized Gumbel PDF's, with arbitrary $m$, 
should probably not be considered more than a a convenient one
parameter fit (for instance, Tracy-Widom distributions defined in
terms of Airy functions \cite{prae}, can be also fitted very well by
generalized Gumbel distributions in the range of interest). 

\section*{Acknowledgements}

We thank G. Giacomin and F. Toninelli for fruitful discussions.

\appendix

\section{Fixman-Freire scheme for the interface model } 

\label{appendixFF}

The Fixman-Freire scheme \cite{Fix_Fre}, invented in the context of
DNA denaturation, allows one to compute melting curves of DNA sequences
in a CPU time of order $O(L)$ instead of $O(L^2)$. Here we briefly
explain how to adapt the Fixman-Freire method to the interface model.

The first step consists in replacing the factor $1/(\alpha-\alpha'-1)^{3/2}$
appearing in the recursion relations (Eq. \ref{rec2}) by
\begin{equation}
\label{FF}
\frac{1 }{(\alpha-\alpha'-1)^{3/2}} \simeq \sum_{i=1}^I a_i \
e^{-b_i (\alpha-\alpha'-1)} 
\end{equation}
where the number ${I}$ of couples $(a_{i},b_{i})$ depends
on the desired accuracy. The parameters $(a_{i},b_{i})$ are determined by a
set of non-linear equations.
This procedure has been tested on DNA chains of length up to $L=10^6$
base pairs \cite{Meltsim,Yer}, and the choice ${I}=15$ gives an accuracy   
better than $0.3\%$ and we have adopted this value.
The table of parameters $(a_{i},b_{i})$ we have used
can be found in our previous work on the wetting transition
\cite{wetting2005}. 

Let us now introduce the following notations
\begin{eqnarray}
R^{\pm}(\alpha) && \equiv \frac{{\tilde Z}^{\pm}(\alpha)}{{\tilde
Z}(\alpha)} \\ 
\frac{1}{S(\alpha)} && \equiv \frac{{\tilde Z}(\alpha)}{{\tilde Z}(\alpha)}
= R^{+}(\alpha)+R^{-}(\alpha)
\label{defrpm}
\end{eqnarray} 
as well as the coefficients $\mu_j^{\pm}(\alpha)$ defined in terms of the Fixman-Freire coefficients
$(b_j)$ (Eq. \ref{FF}) where $j=1,2, ... I=15$ 
\begin{eqnarray}
\mu^+_j(\alpha-2) && =
e^{-b_j (\alpha-4)} \sum_{\alpha'=1,3,..}^{\alpha-4} e^{b_j \alpha'} 
S(\alpha-2) S(\alpha-4) ... S(\alpha'+2) \\
\mu_j^-(\alpha-2) && =
e^{-b_j (\alpha-4)-2 \beta V(\alpha-4) } \sum_{\alpha'=1,3,..}^{\alpha-4}
 e^{b_j \alpha'+ 2 \beta V(\alpha')} 
S(\alpha-2) S(\alpha-4) ... S(\alpha'+2)
\end{eqnarray} 
It is then easy to obtain that for any $j$, these coefficients 
satisfy the simple recursions
\begin{eqnarray}
\mu^+_j(\alpha) && =
 S(\alpha) \left[ 1+ e^{-2 b_j} \mu^+_j(\alpha-2) \right] \\
\mu_j^-(\alpha) && =
 S(\alpha) \left[ 1+ e^{-2 b_j-2 \beta q_{\alpha-3} } \mu^-_j(\alpha-2) \right]
\label{reccoefsFF}
\end{eqnarray} 
In terms of these coefficients, the recursions  (Eq. \ref{rec2})
become for 
the ratios $R^{\pm}(\alpha)$ (Eq. \ref{defrpm})
 \begin{eqnarray}
R^{+}(\alpha) && =  \sigma \left[ 1+
 \sum_{j=1}^I a_j e^{-3 b_j} \mu^+_j(\alpha-2)  \right]  \nonumber \\
R^{-}(\alpha) && =  \sigma e^{-2 \beta q_{\alpha-1} } \left[ 1+
e^{-2 \beta q_{\alpha-3} } \sum_{j=1}^I a_j e^{-3 b_j} \mu^-_j(\alpha-2)  \right]
\label{rec3}
\end{eqnarray}

The recursions (\ref{reccoefsFF},\ref{rec3}) can now be iterated
in a CPU time of order $O(L)$.


\begin{thebibliography}{99}

\bibitem{review1}
T. Garel, H. Orland and E. Pitard, ``Protein folding and
heteropolymers'', in {\em Spin Glasses
and Random Fields}, A.P. Young (ed.), World Scientific, Singapore
(1997) p. 387-443.


\bibitem{Fleer} G.J. Fleer, M.A. Cohen-Stuart, J.M.H.M. Scheutjens,
T. Cosgrove and B. Vincent, {\it Polymers at interfaces}, Chapman and
Hall (1994).

\bibitem{protein1} {\it Proteins at liquid interfaces}, D. M\"obius
and R. Miller Eds, Studies in Interface Science vol. {\bf 7}, Elsevier
(1998).

\bibitem{garel} 
 T. Garel, D.A. Huse, L. Leibler and H. Orland,
Europhys. Lett. {\bf 8}, 9 (1989).


\bibitem{experiments} 
H.R. Brown, V.R. Deline and P.F. Green,
Nature {\bf 341} , 221 (1989);
C.A. Dai, B.J. Dair, K.H. Dai, C.K. Ober, E.J. Kramer, C.Y. Hui and
L.W. Jelinsky, Phys. Rev. Lett. {\bf 73} , 2472 (1994).

 \bibitem{yeung} 
 C. Yeung, A.C. Balazs and D. Jasnow,
Macromolecules, {\bf 25} , 1357 (1992).

\bibitem{sommer}
J.U. Sommer, G. Peng and A. Blumen,
Phys. Rev. E {\bf 53} 5509 (1996);
J. Phys. II France {\bf 6} 1061 (1996);
J. Chem. Phys. {\bf 105} 8376 (1996). 

\bibitem{stepanow}
S. Stepanow, J.U. Sommer, and I.Y. Erukhimovich,
Phys. Rev. Lett. {\bf 81} 4412 (1998).

\bibitem{maritanreplica}
A. Trovato and A. Maritan, Europhys. Lett. {\bf 46}, 301 (1999)


\bibitem{maritanbounds}
 A. Maritan, M.P. Riva and A. Trovato, J.Phys. A, {\bf 32}, L275
(1999). 

\bibitem{Gan_Bre}
V. Ganesan and H. Brenner, Europhys. Lett., {\bf 46}, 43 (1999).

\bibitem{Chen} Z. Y. Chen, J. Chem. Phys., {\bf 111}, 5603 (1999);
ibid., {\bf 112}, 8665 (2000); ibid. {\bf 113}, 10377 (2000).

\bibitem{CM2000} C. Monthus, Eur.Phys. J, {\bf 13}, 111 (2000).

 \bibitem{sinai}
Y.G. Sinai, Theor. Prob. Appl. {\bf 38} , 382 (1993).

 \bibitem{albeverio}
S. Albeverio and X.Y. Zhou, 
J. Stat. Phys. {\bf 53} , 573 (1996).

\bibitem{bolthausen}
E. Bolthausen and F. den Hollander, 
Ann. Prob. {\bf 25} , 1334 (1997).

\bibitem{biskup}
M. Biskup and F. den Hollander, 
Ann. Appl. Prob., {\bf 9}, 668 (1999).

\bibitem{GT1} T. Bodineau and G. Giacomin, J. Stat. Phys. {\bf 117},
801 (2004).

\bibitem{GT2} G. Giacomin and F.L. Toninelli, math.PR/0502394,
Probab. Theor. Rel. Fields (Online, 2005).

\bibitem{GT3} F. Caravenna, G. Giacomin and M. Gubinelli, preprint,
to appear in J. Stat. Phys.

\bibitem{GT4} G. Giacomin and F.L. Toninelli, math.PR/0510047

\bibitem{domany95}
S. Wiseman and E. Domany,
Phys Rev E {\bf 52}, 3469 (1995).

\bibitem{AH}
A. Aharony, A.B. Harris,
Phys Rev Lett {\bf 77}, 3700 (1996).

\bibitem{Paz1} F. P\'azm\'andi, R.T. Scalettar and G.T. Zim\'anyi,
Phys. Rev. Lett. {\bf 79}, 5130 (1997).

\bibitem{domany}
S. Wiseman and E. Domany,
Phys. Rev. Lett. {\bf 81}, 22 (1998) ; Phys Rev E {\bf 58}, 2938 (1998).

\bibitem{AHW}
A. Aharony, A.B. Harris and S. Wiseman,
Phys. Rev. Lett. {\bf 81}, 252 (1998).

\bibitem{Paz2} K. Bernardet, F. P\'azm\'andi and G.G. Batrouni,
Phys. Rev. Lett. {\bf 80}, 4477 (2000).

\bibitem{Fix_Fre} M. Fixman and J.J. Freire, Biopolymers, \textbf{16}, 2693
(1977)

\bibitem{PSdisorder} C. Monthus and T. Garel, cond-mat/0509479.

\bibitem{wetting2005} T. Garel and C. Monthus, Eur. Phys. J. B, {\bf
46}, 117 (2005). 

\bibitem{Bramwell}
S.T. Bramwell et al., Phys. Rev. Lett., {\bf 84}, 3744 (2000);
Phys. Rev. E, {\bf 63}, 041106 (2001).

\bibitem{BBW}
B.A. Berg, A. Billoire and W. Janke, Phys. Rev. E, {\bf 65}, 045102
(2002).

\bibitem{Palassini}
M. Palassini,  cond-mat/0307713 and references therein.

\bibitem{Rosso}
C.J. Bolech and A. Rosso, Phys. Rev. Lett., {\bf 93}, 125701 (2004).

\bibitem{Bertin}
E. Bertin, Phys. Rev. Lett., {\bf 95}, 170601 (2005) and references
therein. 

\bibitem{prae}
M. Pr\"ahoher and H. Spohn, http://www-m5.ma.tum.de/KPZ/.

\bibitem{Meltsim} R.D. Blake, J.W. Bizarro, J.D. Blake, G.R. Day, S.G.
Delcourt, J. Knowles, K.A. Marx and J. SantaLucia Jr., Bioinformatics, 
\textbf{15}, 370-375 (1999); see also J. Santalucia Jr., Proc. Natl. Acad.
Sci. USA, \textbf{95}, 1460 (1998).

\bibitem{Yer} E. Yeramian, Genes, \textbf{255}, 139, 151 (2000).


\end{thebibliography}
\end{document}